\documentclass[preprint,figures]{epl}

\usepackage{cite}
\usepackage{amsmath,amsfonts,bm}
\usepackage{psfrag}

\title{Polymers in linear shear flow: a numerical study}
\shorttitle{Polymers in linear shear flow}
\author{A. Celani \inst{1} \and A. Puliafito\inst{1,3} \and K. Turitsyn\inst{1,2,3}}
\shortauthor{A. Celani \etal}
\institute{
\inst{1} Institut Non Lin\'eaire de Nice - UMR 6618 CNRS, 1361, Route des Lucioles, 06560 Valbonne, France\\
\inst{2} Landau Institute for Theoretical Physics - Kosygina 2, Moscow 119334, Russia \\
\inst{3} CNLS, Theoretical Division, Los Alamos National Laboratory, Los Alamos, NM 87545, USA
}

\begin{document}

\maketitle

\begin{abstract}
We study the dynamics of a single polymer subject to thermal fluctuations
in a linear shear flow. The polymer is modeled as a finitely extendable 
nonlinear elastic (\acro{FENE}) dumbbell.
Both orientation and 
elongation dynamics are investigated numerically as a function of the shear
strength, by means of a new efficient integration algorithm.
The results are in agreement with recent
experiments.
\end{abstract}

\section{Introduction}
\begin{figure}
\psfrag{R}{$\vec{R}$}
\psfrag{x}{x}
\psfrag{y}{y}
\psfrag{z}{z}
\psfrag{f}{$\phi$}
\psfrag{q}{$\theta$}
\onefigure[width=7cm]{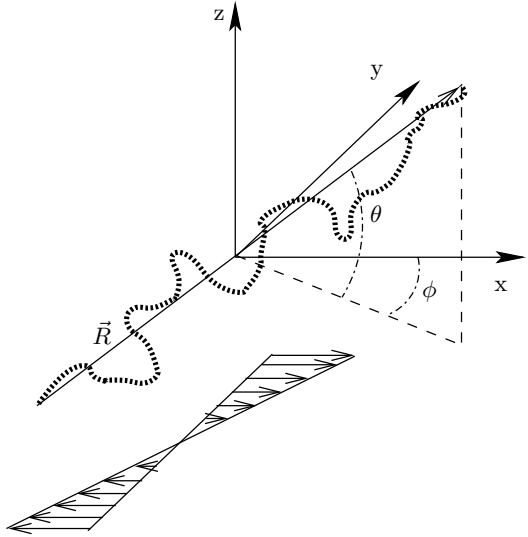}
\caption{A sketch of the geometry of polymer motion in a linear shear flow.}
\label{fig:tumbgeo}
\end{figure}

\begin{figure}
  \psfrag{a}{\large A}
  \psfrag{b}{\large B}
\onefigure[width=4cm]{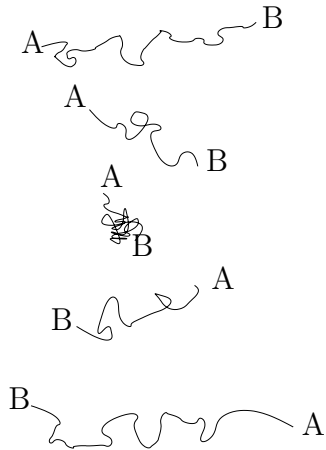}
\caption{Four possible stages of a tumbling event.}
\label{fig:tumbling}
\end{figure}

Nowadays, thanks to the development of effective experimental techniques, 
it is possible to 
follow the motion of a single macromolecule in a flow, either laminar or 
turbulent~\cite{chu,04GCS,05GS,97PSC,98SC,99SBC,01HSBSC,00LLS,97DSG,00HSL,manneville,ladoux,cui,hegner,yin,wuite}. 
This is of crucial importance for applications in polymer processing~\cite{book} and
biophysics~\cite{chu}. Dynamical properties of 
biomolecules have been explored in detail (see 
e.g.~\cite{chu,04GCS,05GS,97PSC,98SC,99SBC,01HSBSC,manneville,ladoux} for DNA
and~\cite{cui} for chromatin) and protein--macromolecule interactions 
have been studied~\cite{hegner,yin,wuite}.\\
The formulation of theoretical models (see e.g.~\cite{book}) able to reproduce 
qualitatively and quantitatively these measurements represents an important
step towards the understanding of single-molecule biophysics.
An extensive analysis of single 
polymer dynamics in simple flows has been conducted in a series of papers by Chu and
coworkers, Larson and coworkers and Shaqfeh and coworkers (see 
\cite{chu,97PSC,98SC,99SBC,01HSBSC,00LLS,97DSG,00HSL} and references therein).
Here we mention in particular two recent papers where the statistics of
orientation and conformation of long-chain molecules in 
linear shear flows has been studied in great detail,
with a direct comparison with numerical models \cite{STSC05,TBSC05}.
An intrinsic difficulty is represented by the large number
of degrees of freedom required to describe the polymer conformation,
and thus its dynamics. Nonetheless, nontrivial aspects of polymer--fluid
interactions may be accounted for and even explained at a 
semi-quantitative level by means of simple, few degrees of freedom models.
One of the simplest, yet reliable, model 
is the finitely extendable nonlinear elastic
dumbbell (FENE)~\cite{book}. The polymer is 
described by its end-to-end distance vector $\mathbf{R}$
and the microphysical properties are essentially 
dumped into two parameters: $1/\gamma$, 
the longest elastic relaxation time of the 
macromolecule, and $\zeta$, its friction
coefficient with the surrounding solvent. 
Nonlinear elastic effects must be accounted for whenever the 
polymer is considerably stretched, as is the case
of shear flows~\cite{99SBC}. 
The geometry of this problem is depicted in Fig.~\ref{fig:tumbgeo}.
The polymer spends a large fraction of time in elongated configurations
along the shear direction. In the following we will present
numerical results about end-to-end orientation, elongation
and about the statistics of tumbling times. The latter is defined
as the time spent between two successive ``flips'' of the polymer ends
(see Fig.~\ref{fig:tumbling}). 
Tumbling can occur via different pathways, e.g. passing by a coiled
state or through folded configurations: those details cannot
be addressed within the single-dumbbell model and will be the subject of
future study. 
\section{From experiments to numerics}
Single polymer orientation and tumbling dynamics was recently studied experimentally by 
Steinberg \etal~\cite{05GS} with a $10^{-3}$\un{ppm} solution of $\lambda-$DNA molecules 
labeled with fluorescence methods (but see also the experiments  
presented in \cite{STSC05,TBSC05}). 
To resolve the angular dynamics two different flow 
configurations were used: one was generated by two discs one of which was rotating 
with uniform angular velocity $\Omega$ and the other flow was 
obtained by a boundary layer in a micro-channel produced with the soft lithography 
method (see~\cite{05GS} and references therein for details).\\
To reproduce the physical situation of~\cite{05GS} we studied a FENE dumbbell in a simple
shear flow $\vec{v}=(sy,0,0)$, sensing thermal fluctuations (see Fig.~\ref{fig:tumbgeo}).
 The equation describing the evolution 
of the end-to-end vector of the polymer is:
\begin{gather}
\dot{R}_i=s\,\delta_{ix}\,R_y-\frac{\gamma R_i}{2} (1-R^2/R_m^2)^{-1} +
\sqrt{\gamma R_0^2}\,\eta_i(t)
  \label{eq:main}
\end{gather}
where $R_0=K_B T/H$, $\gamma=4H/\zeta$,
$\bm{\eta}$ is a three-dimensional white noise with zero mean and correlation
$\langle\eta_i(t)\eta_j(t')\rangle=\delta_{ij}\delta(t-t')$,
$R_m$ is the maximum length of the polymer, $K_B$ is the Boltzmann constant, 
$T$ is the temperature, $\zeta$ is the isotropic drag coefficient, 
and $H$ the spring constant.\\
Even if the single-FENE-dumbbell model does not reproduce precisely the behavior of real 
molecules~\cite{00HSL}, we can set the parameters of our model, $R_0$, $R_m$, 
$\gamma$, as close as possible to their corresponding experimental 
values~\cite{99SBC,05GS}: we choose $R_0\simeq 1\un{\mu m}$, $R_m/R_0 \simeq 21$,
$\gamma\simeq 0.01 \un{s^{-1}}\div 1 \un{s^{-1}}$.\\
The orientation dynamics has been investigated for rigid spheroid by Hinch and 
Leal~\cite{72HL}. As for polymers,at large Weissenberg numbers $\mathrm{Wi}=s/\gamma$, 
where $s$ is the shear rate, the basic ingredients of the polymer dynamics can be 
summarized as follows~\cite{72HL,89Liu,97DSG,04CKLTa,04CKLTb,05T,05PT}:
due to the shear flow the polymer tends to reach the unstable equilibrium
configuration where it is fully extended along the shear direction. In polar 
coordinates $(R,\theta,\phi)=(R_m,0,0 \mbox{ or } \pi)$. 
The effect of thermal noise is to drive the polymer away from this configuration. 
The most probable value of $\theta$ is zero,
due to the symmetry of the dynamics along the $z$ axis.
However, large fluctuations in the off-shear-plane angle can occur.
The most probable value for $\phi$ will be slightly larger than $0$, or $\pi$:
the symmetry-breaking effect of shear causes the polymers to ``hesitate''
for some time before crossing the $x$ axis and then give rise to a 
tumbling event. Few results can be obtained analytically 
for this model, except for the linear case where $R_m/R_0 \to \infty$. 
\section{Numerical algorithm}\label{sec:numerics}
Several numerical methods have been proposed to simulate polymer dynamics 
(see for example~\cite{oett}). 
A commonly encountered problem with nonlinear elastic models is the loss of 
accuracy 
close to the singularity $R\to R_m$. In order to overcome this problem it is possible
to perform a change of variables in the vicinity of $R_m$ that
removes the singularity and allows to use a straightforward time-marching 
scheme. This method can be easily extended 
to other nonlinear models~\cite{91MS} as well as to other flows.\\ 
Eq.~\eqref{eq:main} can be solved by any stochastic discretization
scheme (Euler-It\^o in our case) in the region $R< R_{thr}$, where 
$R_{thr}$ is a fraction of $R_m$, say $0.5 R_m$. Whenever $R$ exceeds 
the threshold we switch to polar variables $(R,\hat{\bm{n}})$, 
where $\hat{\bm{n}}$ is the unity vector describing the orientation of the polymer 
$\hat{n}_i=R_i/R$, and then to the new variables $(z,\hat{{\bm n}})$, where
\begin{equation}
 z=-\frac{R_m}{2}\left(1-\frac{R}{R_m}\right)^2\, .
\end{equation}
This relation can be easily inverted to give $R$ as a function of $z$.
After computing all the contact terms in the It\^o 
convention we have the following equations for $(z,\hat{{\bm n}})$: 
\begin{gather}
  \dot{z}=-\gamma R(1+\frac{R}{R_m})^{-1}+s \hat{n}_x \hat{n}_y R(1-\frac{R}{R_m})+
\frac{\gamma R_0^2}{2}\left(\frac{2}{R}-\frac{1}{R_m}\right)+
\sqrt{\gamma R_0^2}(1-\frac{R}{R_m})\hat{n}_i\eta_i(t)\\
\partial_t \hat{n}_i=s(\hat{n}_y - \hat{n}_y \hat{n}_x^2)\delta_{ix}
-\frac{\gamma R_0^2}{R^2}\big(1-\frac{R}{R_m}\big)^2 \hat{n}_i+
\frac{\sqrt{\gamma R_0^2}}{R}\big(\eta_i(t)-\eta_j(t)\hat{n}_j \hat{n}_i\big)\,.
\end{gather}
which is regular in the neighborhood of $R=R_m$, i.e. $z=0$.

\section{Results}\label{sec:results}
\begin{figure}
  \onefigure[width=7cm]{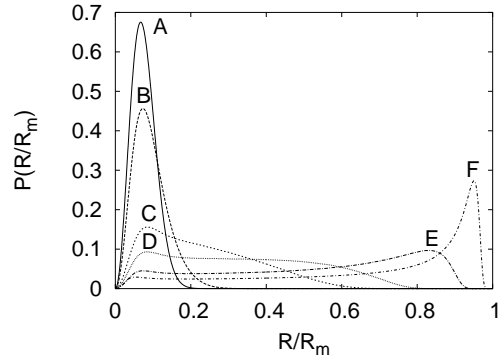}
\caption{The PDF of elongation plotted for 
$\mathrm{Wi}=0,1,5,10,40,200$ (from A to F).}
\label{fig:pdr}
\end{figure}

The Probability Density Function (PDF) of
the modulus of the conformation vector depends strongly on  $\mathrm{Wi}$.
At sufficiently small $\mathrm{Wi}\approx 1$ the statistics does not differ much from
the linear elastic case, since $R \ll R_m$. 
The PDF in the FENE case can be computed analytically only in asymptotic 
regimes~\cite{05VA,05PT}.
The numerical result is shown in Fig.~\ref{fig:pdr} for several
Weissenberg numbers.
At very large $\mathrm{Wi}$ the elongation PDF presents a peak 
with height  scaling as $\mathrm{Wi}^{2/3}$ and width as $\mathrm{Wi}^{-2/3}$ 
(not shown). This result is in agreement with the predictions of Ref.~\cite{04CKLTb}.
As a side remark we notice that experimental measurements of the elongation PDF 
at $\mathrm{Wi}$ as large as $\mathrm{Wi}=76$ do not display a peak near $R_m$ (see 
\cite{00HSL} and fig.~5 of~\cite{99SBC}) as well as numerical results of 
multi-beads models (see figs.~4, 5
in \cite{00HSL} and the discussion therein).
 
\begin{figure}
  \twofigures[width=7cm]{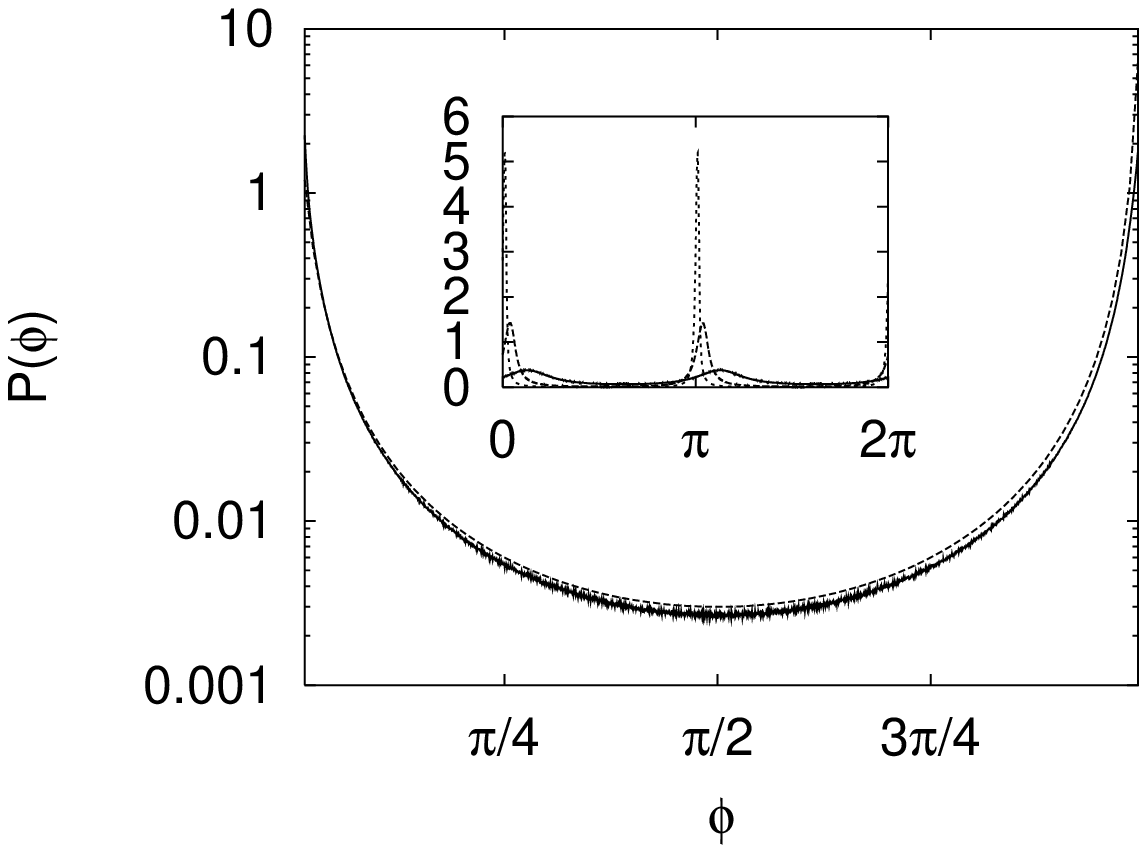}{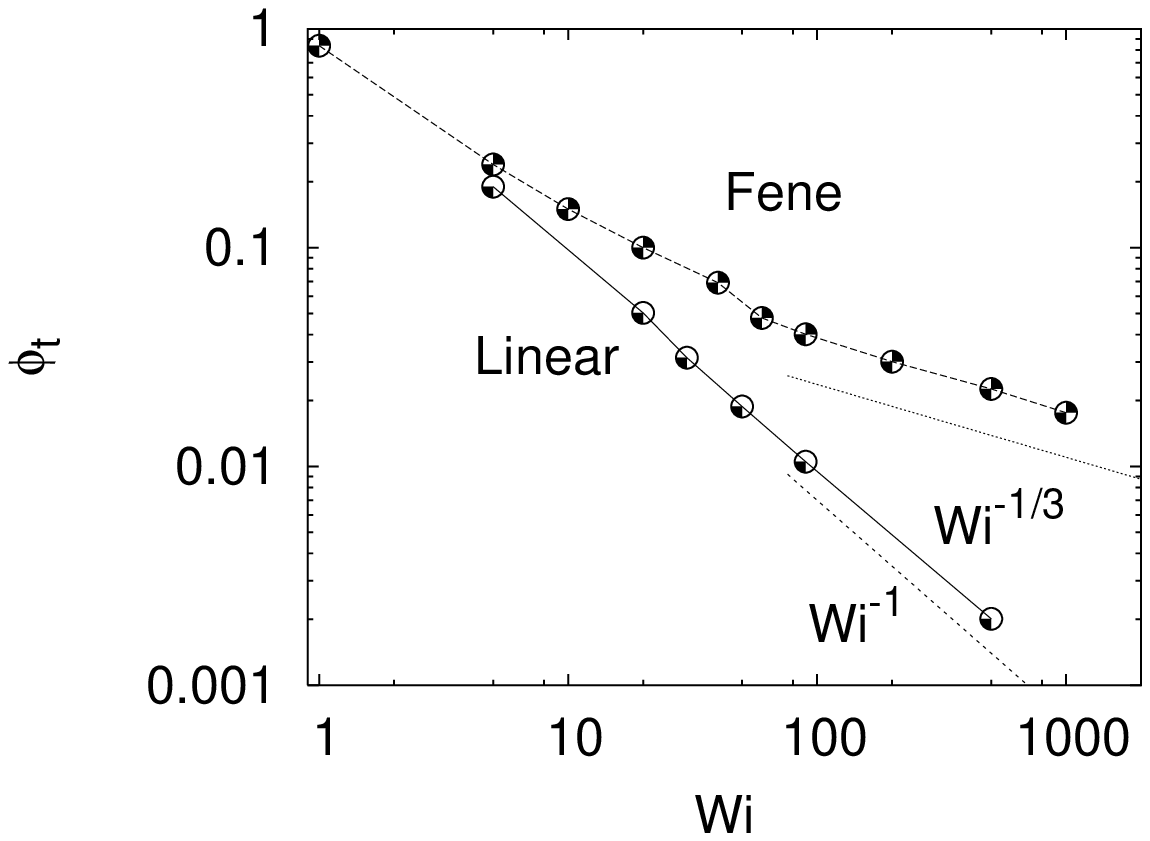}
\caption{The PDF of the angle $\phi$ plotted with $\sin^{-2}{\phi}$. In the inset 
the PDF in linear scale plotted for $\mathrm{Wi=1,5,40}$.}
\label{fig:pdphi}
\caption{The behavior of $\phi_t$ as a function of $\mathrm{Wi}$.}
\label{fig:asyphi}
\end{figure}

The orientation of polymers follows the qualitative picture drawn in the linear elastic 
case~\cite{05T,05PT}, even though there appear important quantitative differences. 
The PDF of the \mbox{in-shear-plane angle} $\phi$ is shown
in Fig.~\ref{fig:pdphi}. The probability is concentrated in the vicinity of 
$\phi=0,\pi$  with a peak width at half height $\phi_t$, whose dependency 
on $\mathrm{Wi}$ is shown in Fig.~\ref{fig:asyphi}. The case of a linear elastic
dumbbell $R_m=\infty$ is shown for comparison.
The angle $\phi_t$ decreases with $\mathrm{Wi}$ in both cases, 
i.e. the larger is $\mathrm{Wi}$ the 
narrower is the region around the $x$ axis where the 
polymer spends most of its time. 
The scaling can be derived by simple physical arguments: following Chertkov 
\etal~\cite{04CKLTa} the evolution equation for $\phi$ in the region 
$\phi\ll1$ is approximated by
by $\partial_t \phi=-s\phi^2+\sqrt{\gamma R_0^2/R^2}\,\eta_\phi$, where $\eta_\phi$ is a 
white noise. Thus $\phi_t$ can be estimated balancing 
shear and noise terms in the right hand side terms, i.e. 
$\phi_t\sim \mathrm{Wi}^{-1/3}(R_0/R)^{2/3}$. At large $\mathrm{Wi}$, 
for a linear elastic dumbbell
one has $R \propto \mathrm{Wi}$, yielding $\phi_t \sim \mathrm{Wi}^{-1}$, whereas for a
nonlinear elastic force one estimates $R \sim R_m$ to find 
$\phi_t \sim \mathrm{Wi}^{-1/3}$. The tails of the PDF follow closely the distribution 
$\sin^{-2}\phi$ dictated by the shear (see Fig.~\ref{fig:pdphi}).\\
The agreement with the experiments is very good~\cite{05GS}: the scaling 
in the tail follows $\sin^{-2}{\phi}$, and the dependence of $\phi_t$ on 
$\mathrm{Wi}$ is close to the theoretical prediction already for $\mathrm{Wi}=25$. \\

\begin{figure}
  \twofigures[width=7cm]{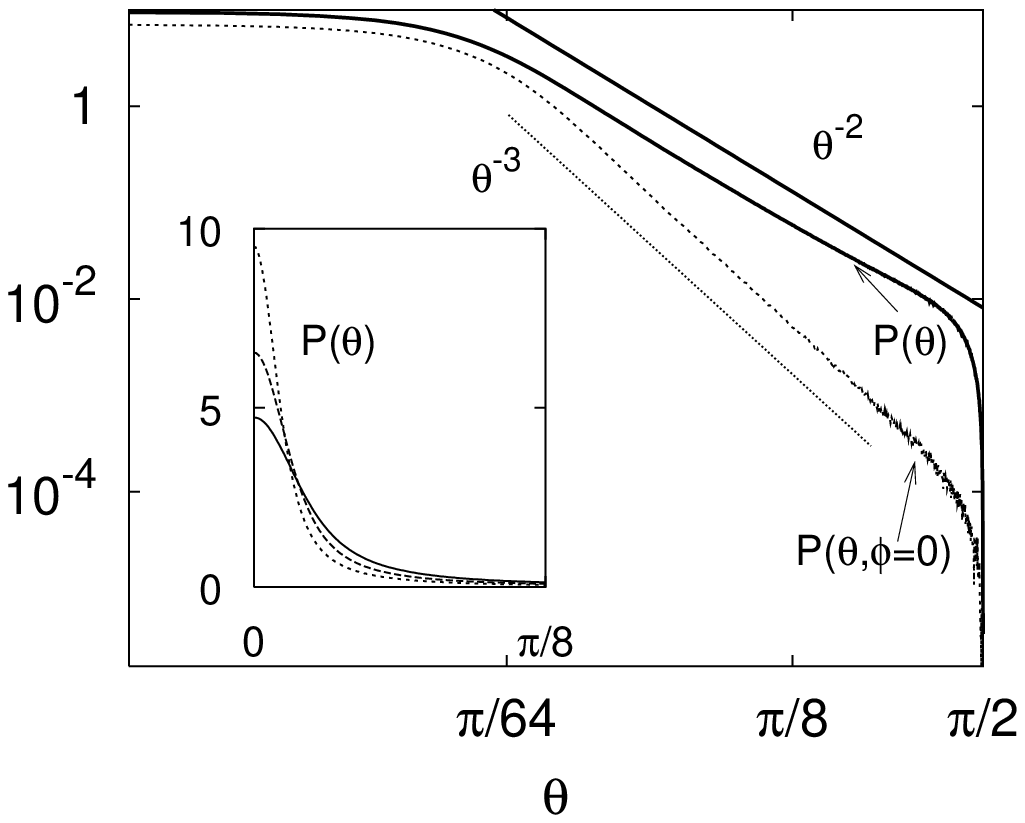}{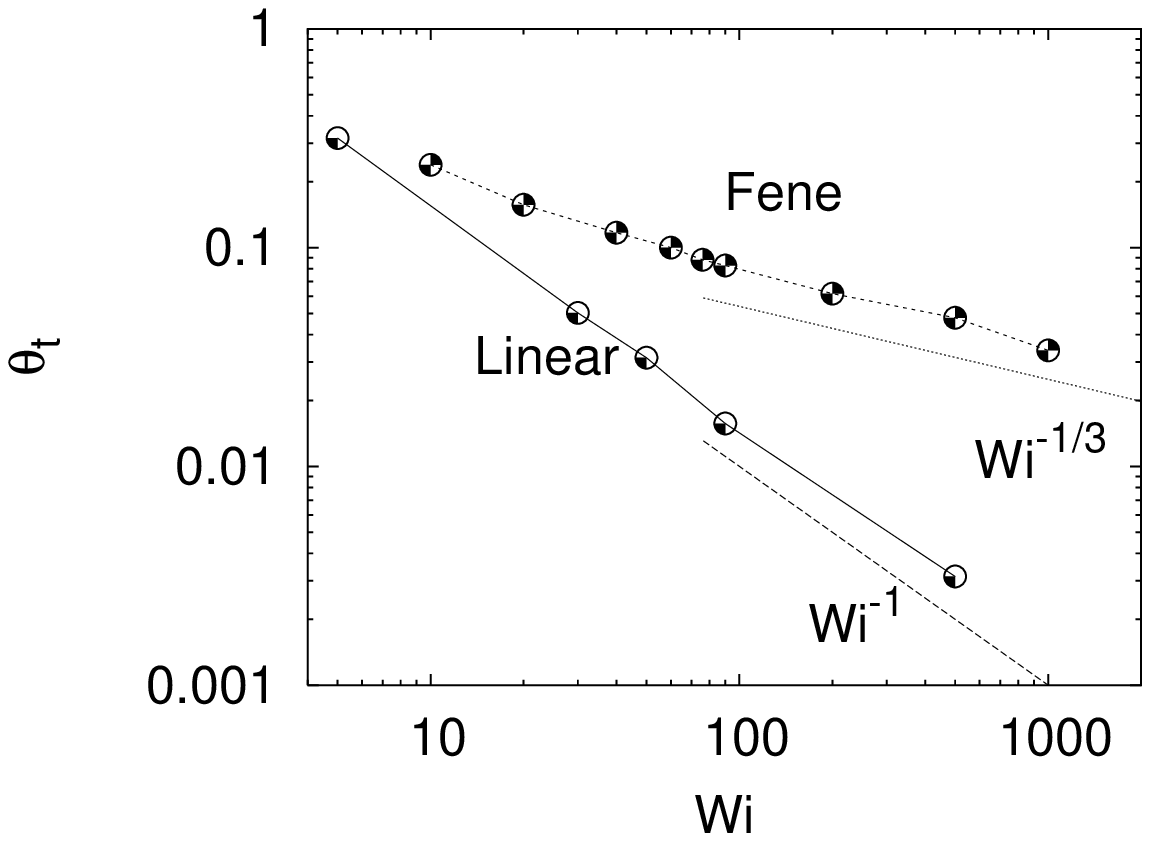}
\caption{ The PDF of the angle $\theta$ plotted against 
$\theta^{-2}$. In the inset the PDF in linear scale 
for $\mathrm{Wi}=20,40,100$.}
\label{fig:pdtheta}
\caption{The behavior of $\theta_t$ as a function of $\mathrm{Wi}$.}
\label{fig:asytheta}
\end{figure}

The marginal PDF of the angle $\theta$ is presented in Fig.~\ref{fig:pdtheta}. 
The tails decay as $\theta^{-2}$, with a scaling range increasing
with $\mathrm{Wi}$. The algebraic behavior has been observed in~\cite{05GS} for 
$\mathrm{Wi}=17.6$, and even if $\mathrm{Wi}$ is not very high 
the agreement is remarkable. The probability density of $\theta$ for 
 small angles $\phi \sim \phi_t$, or equivalently the joint
 PDF $P(\theta,\phi=0)$, shows a neat power law close to $\theta^{-3}$ 
 for $\theta \gg \theta_t$. This nontrivial scaling behaviour has been
 predicted theoretically and observed numerically for the linear elastic case 
 in Refs.~\cite{04CKLTa,05T,05PT}. 
The width of the peak of the $P(\theta)$ at half height, 
$\theta_t$, decays as $\mathrm{Wi}^{-1/3}$ for the nonlinear elastic case. The 
agreement with experimental data is perfect~\cite{05GS}.\\ 
Note that the cross-over between the linear elastic case $\phi_t\sim\theta_t\sim 
\mathrm{Wi}^{-1}$ and the FENE case $\phi_t\sim\theta_t\sim \mathrm{Wi}^{-1/3}$ occur 
at $\mathrm{Wi}\sim 1$, as measured experimentally.\\

\begin{figure}
  \twofigures[width=7cm]{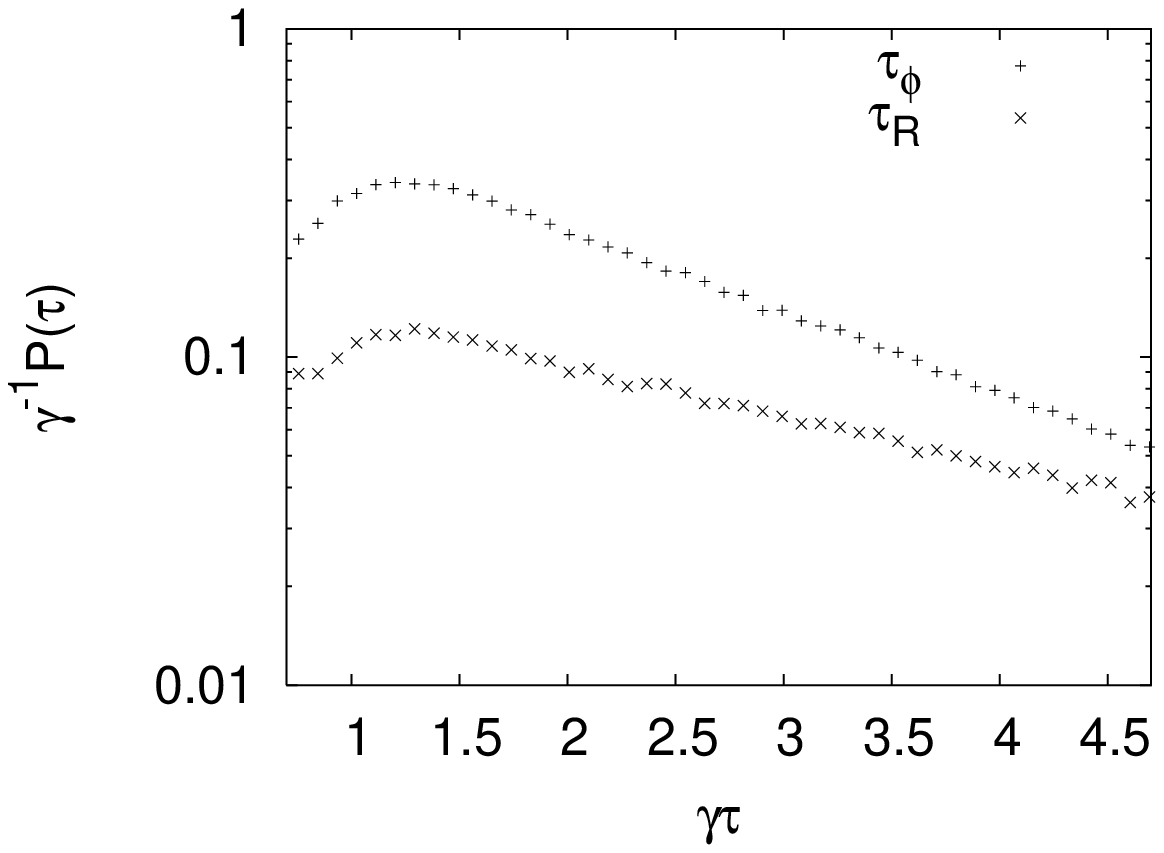}{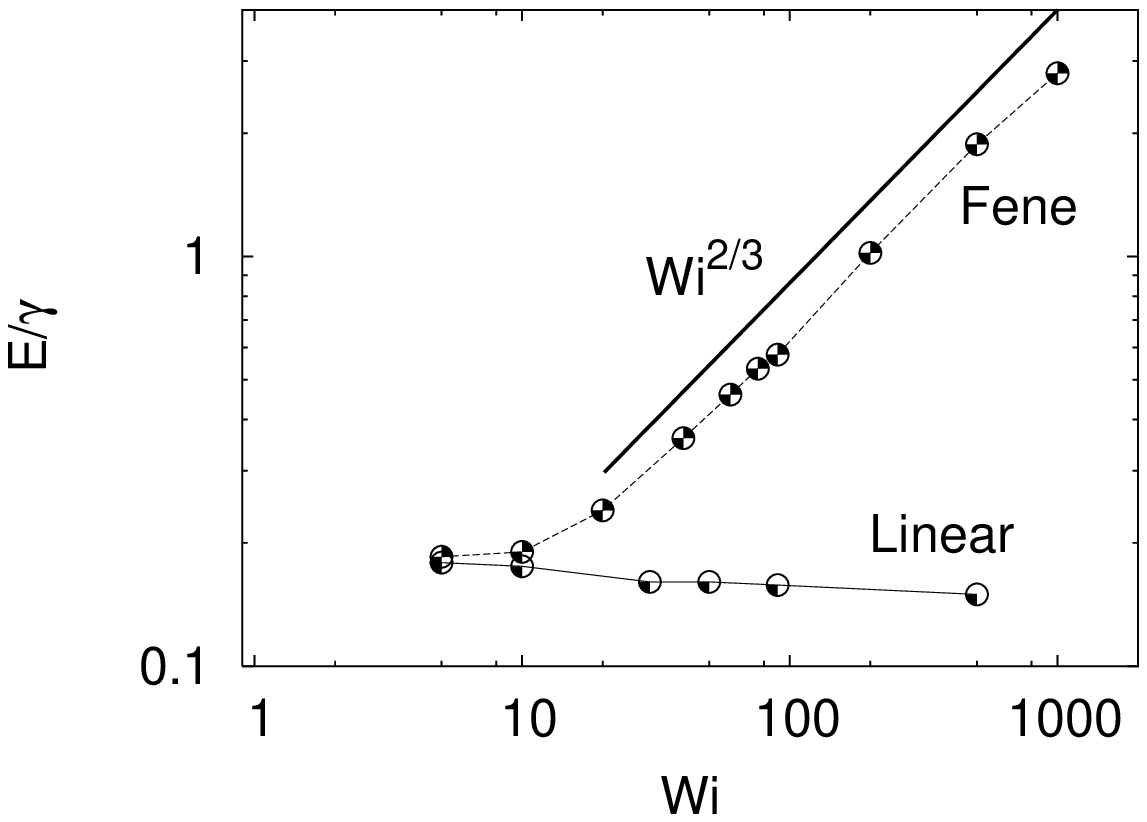}
\caption{The PDF of tumbling times: 
$\tau_R$ is the time elapsed between two nonadjacent coiled states, 
defined as the states 
where $R$ is smaller than $\lambda R_0$ ($\lambda=1.5$ in this figure); 
$\tau_\phi$ 
is defined as the time between two rotations of $\pi$ in the angle $\phi$.
Here, $\mathrm{Wi}=76$.}
\label{fig:pdtau}
\caption{The exponent $E$ rescaled with the relaxation time as a function of 
$\mathrm{Wi}$ in the linear case and in the FENE case.}
\label{fig:asytau}
\end{figure}

For what concerns the tumbling times statistics 
there are two possible definitions~\cite{05PT}: 
{\it(i)\/} given an appropriate threshold value in $R$ that defines the
coiled state for the polymer, one can 
compute the time spent during two successive coiled states~\cite{05GS}.
{\it(ii)\/}
one can consider the time between two subsequent
crossings of the plane $\phi=\pi/2$.\\
Both definitions are ambiguous for small values of $\mathrm{Wi}$,
i.e. when the polymer spends most of its time in a coiled state.
For large tumbling times the PDF is exponential for both definitions of $\tau$,
$P(\tau)\sim \exp{[-E\tau]}$ (see Fig.~\ref{fig:pdtau}). This is a robust  
feature of this phenomenon~\cite{04CKLTa,05GS}. 
Experimental measurements of the tumbling time are possible only following the 
first definition, due to lack of angular resolution~\cite{05GS}. \\
The exponent $E$ of the tail in the linear elastic case is inversely proportional to the
relaxation time of the polymer $\gamma^{-1}$. In the FENE case there is a non trivial 
dependence on $\mathrm{Wi}$, as shown in Fig.~\ref{fig:asytau}. 
The scaling of the typical tumbling time 
$\tau_t \sim E^{-1}$ can be estimated at large $\mathrm{Wi}$
as follows: the angular motion in the region $\phi \sim \phi_t$ is
driven by the thermal noise and is therefore diffusive. The diffusion
coefficient is $D= \gamma (R_0/R)^2$  and therefore $\tau_t \sim \phi_t^2/D$.
Substituting $R\propto R_0 \mathrm{Wi}$ for the linear spring model and $R \sim R_m$
for the FENE model one obtains for $E/\gamma$ the scalings $\mathrm{Wi}^0$ and
$\mathrm{Wi}^{2/3}$, respectively. \\
The behavior of the PDF at $\tau \ll \tau_t$ is model dependent and should not
be considered as relevant (see e.g.~\cite{05PT}). 
In experiments~\cite{05GS} 
the exponential tail of the PDF is observed and 
the dependence of $\tau_t$ on $\mathrm{Wi}$ is in 
accordance with theoretical arguments and numerical results.

\section{Conclusions}
We studied the dynamics of a single FENE polymer immersed in a simple shear flow with 
thermal noise. The statistics of orientation, elongation and tumbling 
of the polymer have been analyzed
in comparison with experimental measurements~\cite{05GS}, previous numerical 
simulations~\cite{00HSL,05PT}, and theoretical expectations \cite{04CKLTa,04CKLTb,05T}. 
Even if the large variety of conformations of real polymer
can not be explored within such a simple model, single-FENE-dumbbell 
can reproduce semiquantitatively several aspects of the behavior of real 
polymers.
\acknowledgments
We acknowledge innumerable inspirating discussions with  M. Chertkov, 
V. Lebedev, V. Steinberg. 
This work was supported by EU network HPRN-CT-2002-00300, and by the PROCOPE grant 
N\textordmasculine 07574. AP and KT acknowledge CNLS at Los Alamos National Laboratory for 
hospitality and support.

\end{document}